\documentstyle[12pt]{article}
\begin{document}
{\it Cavendish Preprint HEP 93/4 \\
     Adelaide University ADP-93-218/T136 \\
     13th October 1993}
\vspace{1.0cm}
\begin{center}
{\huge \bf The Spin Structure of the Nucleon \\}
\vspace{3ex}
\vspace{3ex}
{\large \bf S. D. Bass$^1$ and A. W. Thomas$^2$ \\
\vspace{3ex}
{\it $^{1}$HEP Group, Cavendish Laboratory, \\
University of Cambridge, Madingley Road, Cambridge, CB3 0HE,
England \\}
\vspace{3ex}
{\it $^{2}$Department of Physics and Mathematical Physics,
University of Adelaide, \\
Adelaide, SA 5005, Australia \\ }       }

\vspace{3ex}
{\large \bf Abstract \\}
\end{center}
\vspace{3ex}

{\large

We provide an introduction to the ideas of spin-dependent deep-inelastic
scattering. Recent experimental results are summarised and possible
explanations related to the axial anomaly presented. Further experiments
that could greatly clarify the situation are also described.}

\vspace{3.0cm}

\pagebreak

{\it Polarisation data has often been the graveyard of fashionable theories.
If theorists had their way they might well ban such measurements altogether
out of self-protection.}

J. D. Bjorken, Proc. Adv. Research Workshop on QCD Hadronic Processes,
St. Croix, Virgin Islands (1987).

\vspace{3ex}

\section {\bf Introduction}

Unravelling the mystery of the structure of the nucleon is one
of the great challenges facing modern physics. On the theoretical
side there are a vast array of QCD motivated models as well as the more
fundamental lattice calculations. These are commonly compared with low
energy data such as excitation energies and electoweak form-factors. In
this regard it is somewhat surprising that little use has been made of
deep-inelastic scattering data.

Deep-inelastic scattering (DIS) experiments provide perhaps our cleanest
window on hadron structure at large momentum transfer squared $Q^2$.
The original DIS experiments at SLAC in the 1960s showed that the form
factors in $e p$ scattering exhibit approximate {\it scaling} at large
$Q^2$.
This remarkable observation was celebrated recently when the 1990 Nobel Prize
in Physics was awarded to the experimenters Friedman,
Kendall and Taylor [1].
It gave rise to the original parton ideas of Feynman [2] and Bjorken and
Paschos [3,4].
Precise data revealed the logarithmic scaling violations and also the gluon
distribution that were predicted by QCD.

While the QCD improved parton model has proven extremely effective in
correlating DIS data, it has only recently been taken seriously as
a means to test and refine the models mentioned earlier [5, 6].
The motivation
for doing so has increased dramatically since the measurement of the
spin structure function $g_1^p(x)$ by the European Muon Collaboration
(EMC) in 1988 [7]. That data has been quite widely interpreted as sparking
a crisis -- the ``spin crisis''.

Our purpose here is to introduce the basic ideas of the QCD improved
parton model as it relates specifically to spin-dependent
measurements. We shall review the current experimental situation and why
the data was viewed as such a challenge to the quark model.
We then discuss
the axial anomaly in QCD and explain how it may contribute to polarised DIS.

This review is aimed at the broad audience of experimentalists and
theorists not directly working in this area. It should be particularly
valuable for graduate students interested in the problem. For those who
want to follow the theoretical developments in more detail we refer to
our recent review [8], as well as to the many original papers cited
here. Other reviews with a different emphasis may be found in the
lecture by Veneziano [9]
and in the proceedings of the recent international conference on particles
and nuclei [10].
Those for whom the experimental details are of greatest
interest should refer to the recent review of Windmolders [11].

\section {\bf Deep Inelastic Scattering}

Our primary concern is with deep-inelastic
scattering from polarised targets -- at CERN using muons
(by the European Muon Collaboration (EMC) and the
Spin Muon Collaboration (SMC) ) and at SLAC using electrons.
Consider a DIS experiment where a muon beam with definite
polarisation and momentum
$k^{\mu}=(E;{\vec k})$ scatters from a polarised proton target.
At leading order in the electromagnetic
interaction this is shown in the one photon exchange diagram of Fig. 1.
We work in the laboratory frame so that the proton target
has momentum $p^{\mu}=(m_p;0_{T},0)$ and polarisation
$S^{\mu}$.
The muon is scattered through an angle $\theta$ to emerge
with momentum $k^{' \mu} = (E^{'}; \vec{k}^{'})$.
The exchanged photon carries momentum $q^{\mu} = (k-k^{'})^{\mu}$.
The scattering process is then characterised by the two invariants
$Q^2 = -q^2$ and $\nu = p.q $
($\nu = m_p (E-E^{'}) $ in the LAB frame) or, equivalently, by
$Q^{2}$ and the Bjorken variable
$x = {Q^2 \over 2 \nu}$.
We measure the inclusive hadronic cross section, so that
hadronic final states, $X$, with the same invariant mass squared,
$W^{2}=(p+q)^{2}$,
are not separated.

The differential cross section for the one photon exchange process
(figure 1) is given by [12]
$$
{d^2 \sigma \over d\Omega d E^{'} } =
{ \alpha^2 \over q^4 } {E^{'} \over E} L_{\mu \nu} W^{\mu \nu}.
\eqno(1)
$$
Here, $\alpha$ is the fine structure constant and $L_{\mu \nu}$
and $W_{\mu \nu}$ describe the muon and hadronic vertices respectively.
Since the muon is elementary we can write down an
exact expression for $L_{\mu \nu}$ from the Feynman rules, viz.
$$
L_{\mu \nu} =
2 \Biggl[ k_{\mu} k_{\nu}^{'} + k_{\mu}^{'} k_{\nu}
 - g_{\mu \nu}(k.k^{'} - m_{\mu}^2)
+ i \epsilon_{\mu \nu \rho \sigma} q^{\rho} s^{\sigma} \Biggr],
\eqno(2)
$$
where we take $ \epsilon_{0123} = +1 $.
The hadronic tensor,
$W^{\mu \nu}$, contains all of the information about the hadronic
target that one can extract from such inclusive measurements. Its
form is constrained by symmetry arguments.
We write $W_{\mu \nu}$ as a sum of symmetric and antisymmetric
contributions;
viz. $W_{\mu \nu} = W_{\mu \nu}^S + i W_{\mu \nu}^A$.
Then the requirements of covariance,parity,
charge conjugation, and current conservation
($q^{\mu} W_{\mu \nu} = 0$) imply the form
$$
W_{\mu \nu}^{S} = {1 \over m_p} (-g_{\mu \nu} + {q_{\mu} q_{\nu} \over q^2 })
               F_1 (x, Q^2)
             + {1 \over m_p \nu} (p_{\mu} - {p.q \over q^2} q_{\mu})
             (p_{\nu} - {p.q \over q^2} q_{\nu}) F_2 (x, Q^2)
\eqno(3)
$$
and
$$
W_{\mu \nu}^A = {\epsilon_{\mu \nu \lambda \sigma}q^{\lambda} S^{\sigma}
                \over \nu} g_1 (x, Q^2)
            + {\epsilon_{\mu \nu \lambda \sigma}q^{\lambda}
               (\nu S^{\sigma} - q.S p^{\sigma}) \over \nu^2 }
               g_2 (x, Q^2)
\eqno(4)
$$
The form factors in equs.(3) and (4) contain all of the
target dependent information.

Deep inelastic scattering is defined by the kinematic limit where both
$Q^2 \gg m_p^2$ and $W^2 \gg m_p^2$, so that
we are beyond the resonance region (where $W^2$ may
coincide with the mass squared of one of the excited nucleon
resonances).
In the DIS limit the form factors $F_i$ and $g_i$ in $W_{\mu \nu}$
exhibit approximate scaling.
That is, they behave as structure functions of the single variable $x$ -
modulo a slow logarithmic variation in $Q^2$, which is
described by perturbative QCD.
The $scaling$ property
reveals a local interaction between the hard photon and charged
elementary partons (quarks) inside the proton.
It is the same effect as in Rutherford's $\alpha$ particle
scattering experiments which revealed the nucleus inside the atom [13] -
only at a much deeper level inside the nuclear ``onion" [14].

We now turn to the measurement of the spin dependent
structure function,$g_1$.
This experiment involves a muon (or electron) beam and a proton target,
both with longitudinal polarisation.
We let $\uparrow \downarrow$ denote the longitudinal beam polarisation
and $\Uparrow \Downarrow$ denote (longitudinal) polarisation of
the target proton.
Then the differential cross sections are
$$
{d^2 \sigma \uparrow \Uparrow \over d\Omega dE^{'} } +
{d^2 \sigma \uparrow \Downarrow \over d\Omega dE^{'} } =
{8 \alpha^2 (E^{'})^2 \over m_p Q^4 }
\biggl[ 2  \sin^{2} {\theta \over 2} \ F_1 (x,Q^2) + {m_p^2 \over \nu}
 \cos^{2} {\theta \over 2} \ F_2 (x,Q^2) \biggr]
\eqno(5)
$$
and
$$
{d^2 \sigma \uparrow \Uparrow \over d\Omega dE^{'} } -
{d^2 \sigma \uparrow \Downarrow \over d\Omega dE^{'} } =
{4 \alpha^2 E^{'} \over Q^2 E \nu }
\biggl[ (E+E^{'} \cos \theta ) \ g_1 (x, Q^2) - {2 x m_p}
\ g_2 (x, Q^2) \biggr]
\eqno(6)
$$
One needs both a polarised beam and a polarised target
to measure $g_1$.
The structure function $g_2$ is suppressed with respect to $g_1$
by a factor ${m_p \over E} \sim 0.01$, for a typical (muon)
beam energy of 100GeV, and is lost among the errors.
In this article we shall be interested only in $g_1$.
Readers interested
in the physics of $g_2$ should consult the review by Jaffe [15].

\section {The parton model and sum-rules}

The parton model began with Feynman who showed that the early SLAC
deep inelastic scattering experiments could be explained in terms
of the hard photon scattering incoherently from elementary {\it parton}
constituents in the proton.
The structure functions measure the probability for finding a quark with
momentum fraction $x = {p_+^{parton} \over p_+^{proton}}$
in the proton and which is polarised either in
the same or the opposite direction to the proton's polarisation.
This is usually called the naive parton model; it has no gauge degrees
of freedom.

In the naive parton model the structure functions are described
by the four linearly independent parton distributions.
There are the two spin independent distributions. The charge
conjugation even combination
$$
 (q + {\overline q})^{\uparrow} (x)
  + (q + {\overline q})^{\downarrow} (x)
\eqno(7)
$$
occurs in the structure functions $F_1 (x, Q^2)$ and $F_2 (x, Q^2)$.
It is measured with an unpolarised target
and either an electron or muon beam.
In the naive parton model one finds
$$
F_2 (x) = 2x F_1 (x) = x \sum_q e_q^2
\biggl[ (q + {\overline q})^{\uparrow} (x)
  + (q + {\overline q})^{\downarrow} (x) \biggr],
\eqno(8)
$$
where the relationship between $F_1$ and $F_2$ is called the Callan-Gross
relation.

If the exchanged boson is a $W^{\pm}$ or a $Z^0$ instead of a photon then
there is also a
parity odd, C-odd term in $W_{\mu \nu}$.
One finds a new structure function $F_3$, which may be measured with an
(anti-)neutrino beam and a fixed target
or with an electron beam at HERA, where there is a large centre of mass
energy.
The structure function
$F_3 ^{\nu + {\overline \nu}} (x, Q^2)$ measures the C-odd valence distribution
$$
(q - {\overline q})^{\uparrow} (x)
+ (q - {\overline q})^{\downarrow} (x)
\eqno(9)
$$

The structure functions contain all the information about the target in DIS.
It follows that
when we take a given moment of a structure function
we project out the target matrix element of some gauge-invariant, local
operator.
Indeed, this idea is formalised in the operator product expansion analysis
of deep inelastic scattering.
For example, the first moment of the unpolarised structure function $F_2$
(which is related to the the target matrix element of the energy
momentum tensor)
determines the fraction of the target's momentum carried by the
quarks.
Similarly,
the first moment of the valence distribution tells us the number of valence
quarks in the proton: three!
(This is the Gross Llewellyn Smith sum rule
which has been verified within experimental error.)

The moments of structure functions provide a testing ground for our
understanding of proton structure in QCD.
The first piece of experimental evidence that we needed to go beyond the
naive parton model came when it was observed that quarks only contribute
50\% of the proton's momentum.
This was naturally accomodated with the advent of QCD where the gluons carry
the momentum fraction
$$
<x g(x,Q^2)> \rightarrow {16 \over 16 + 3N_f},
\eqno(10)
$$
in the limit of infinite $Q^2$ [16].
(Here $N_f$ is the number of quark flavours.)

Deep inelastic scattering is continuing to teach us about the QCD structure
of hadrons.
In the last decade we have seen an extensive programme to measure the
structure functions of nuclear targets.
This work was motivated by the discovery of the EMC effect
,namely that the structure
functions of a
nucleon are modified when the nucleon is in a nuclear environment [17, 18].
More recently, the
unpolarised experiments at HERA are providing new information about the
small x region
and, in particular, the nature of the pomeron.
The HERA experiments and E-665 at FNAL are also teaching us about the jet
structure of deep inelastic scattering.

Whilst the physics of unpolarised deep inelastic scattering is reaching
maturity,
we are also seeing a whole new area of physics opening up - the spin dependent
world!
It is more than ten years since the first polarised deep inelastic scattering
experiments were done at SLAC [19].
This work became the subject of renewed interest
5 years ago when EMC [7, 20] extended the
SLAC measurements to smaller x
and announced their
data on the spin dependent proton structure function $g_1$.
The naive parton model interpretation of their data is that the quark partons
contribute very little to the proton's spin -
in contradiction with quark models.

During the last year we been treated to new spin data: the measurements of the
deuteron structure function $g_1^d$ by SMC at CERN [21]
and the $^3He$ structure function by E-142 at SLAC [22]. From
each of these measurements one can extract the neutron
spin structure function $g_1^n$.
Before we discuss the EMC experiment and the new data, let us review what
the naive parton model says about $g_1$.
This way we shall understand why the EMC result created so much interest.

In the naive parton model $g_1$ is written as:
$$
g_1 (x) = {1 \over 2} \sum_q e_q^2 \Delta q(x)
\eqno(11)
$$
where
$$
\Delta q(x) = (q^{\uparrow} + {\overline q}^{\uparrow})(x) -
			(q^{\downarrow} + {\overline q}^{\downarrow})(x)
\eqno(12)
$$
is the polarised quark distribution and $e_q$ denotes the quark charge.
It is helpful to rewrite $g_1(x)$ in terms of the SU(3) flavour
combinations:
$\Delta u(x) - \Delta d(x)$,
$\Delta u(x) + \Delta d(x) - 2 \Delta s(x)$
and
$\Delta u(x) + \Delta d(x) + \Delta s(x)$.

In the naive parton model,
$$
\Delta q = \int_0^1 dx \ \Delta q(x)
\eqno(13)
$$
determines the fraction of the proton's
spin which is carried by quarks (and anti-quarks) of flavour $q$.
Thus, we write
$$
\int_0^1 dx \ g_1 (x) =
         {1 \over 12} (\Delta u - \Delta d)
       + {1 \over 36} (\Delta u + \Delta d - 2 \Delta s)
       + {1 \over 9}  (\Delta u + \Delta d + \Delta s)
       + {2 \over 9}   \Delta c
\eqno(14)
$$
where the singlet term $\Delta u + \Delta d + \Delta s$
denotes the light-quark spin content of the proton.
(Here, we have included a contribution from charm quarks $\Delta c$, which
is present above the charm threshold.)

In operator language, $\Delta q$ is defined by the proton matrix element
of the axial currents.
We write
$$
2m_p S_{\mu} \Delta q_a =
<p,S | {\overline q} \gamma_{\mu} \gamma_5 {\lambda^a \over 2} q |p,S >
\eqno(15)
$$
for $a=3,8,0$.
The non-singlet matrix elements also arise in the neutron and hyperon
beta decays.
Current algebra relates the spin dependent
(strong interaction) structure of the proton
measured im polarised deep inelastic scattering at
high energies to the quantities needed in low energy
weak interaction physics.
The currents which measure $\Delta q_3$ and $\Delta q_8$ are soft operators
so that these quantities are scale independent.
They are determined as
$\Delta q_3 = g_A = F+D$ and
$\Delta q_8 = {1 \over \sqrt{3} }(3F - D)$ within
SU(3). Here $F$ and $D$ are the antisymmetric and symmetric SU(3) couplings.
One finds
$\Delta u - \Delta d = 1.254 \pm 0.006$ and
$\Delta u + \Delta d - 2 \Delta s = 0.688 \pm 0.035$ [23]
independent of scale.
Since $\Delta q_3$ and $\Delta q_8$ are determined from other experiments
we can extract the singlet ``spin content"
of the proton $\Delta u + \Delta d + \Delta s$.
(This assumes that there is a negligible charm component $\Delta c$ in the
data.)

Before we can extract the ``spin content" from $g_1$ data we also need to
allow for the perturbative QCD Wilson coefficients.
Ignoring charm, one finds:
$$
\int_{0}^{1}dx \ g_1^p(x,Q^{2}) = {1 \over 6}
 \biggl[ (\Delta q_3 + {1 \over \sqrt{3}} \Delta q_8)
 (1- {\alpha_s \over \pi} )
 + 2 \sqrt{2 \over 3}  \Delta q_0
 (1- {\alpha_s \over \pi} {33 - 8N_f \over 33 - 2N_f })
 \biggr]
\eqno(16)
$$
where the perturbative coefficients are quoted to $O(\alpha_s)$ [24].
In QCD the singlet term
$\Delta q_0 = {1\over 2} \sqrt{2 \over 3} (\Delta u + \Delta d + \Delta s)$
has an anomalous dimension, due to the axial anomaly,
which starts at two loops [25].

There are two sum-rules for $g_1$, which have been tested in recent data.
The Bjorken sum-rule [3] gives a relation for the difference between the
first moment of $g_1$ for a proton and neutron target.
To order $\alpha_s^3$ precision it reads:
$$
\int_0^1 dx \ \biggl( g_1^p(x) - g_1^n(x) \biggr)
= {1 \over 6} g_A^3
\biggl[ 1 - {\alpha_{s} \over \pi} + C_2 ({\alpha_s \over \pi})^2
+C_3 ({\alpha_s \over \pi})^3 \biggr],
\eqno(17)
$$
where $C_2 = -3.5833$ and $C_3 = -20.2153$ in the three flavour theory [26].
The Bjorken sum-rule was derived using current algebra before the advent of
QCD and is a test of isospin.

The Ellis-Jaffe sum-rule [27] is a test of Zweig's rule
(or OZI after Okuba, Zweig and Iizuka) in the flavour singlet
channel.
If we assume that strange (and heavy) quarks do not play a significant role
and set $\Delta s=0$,
then the quark ``spin content" would be determined by the hyperon beta decays.
In this scenario we would have
$\Delta q_{0} = 0.69 \pm 0.03$,
where the rest of the proton's spin would be carried by the gluons and
also by quark
and gluon orbital angular momentum.
Substituting this number into equ.(16) we find
$\int_0^1 dx \ g_1(x) = 0.189 \pm 0.005$ (at $10 GeV^2$).
The Ellis-Jaffe sum-rule involves a model dependent
assumption of good OZI , whereas
the Bjorken sum-rule should hold exactly in QCD.

We now discuss the experimental results in polarised deep inelastic scattering
and see how these sum-rules hold up.

\section {Gifts from experiment}

The European Muon Collaboration (EMC) measured the asymmetry
$$
A(x,Q^2) =
{ {d^2 \sigma \uparrow \Uparrow \over d\Omega dE^{'} } -
{d^2 \sigma \uparrow \Downarrow \over d\Omega dE^{'} }
\over
{d^2 \sigma \uparrow \Uparrow \over d\Omega dE^{'} } +
{d^2 \sigma \uparrow \Downarrow \over d\Omega dE^{'} } }
\eqno(18)
$$
in the kinematic range
$0.01 < x < 0.7$ and 1.5 GeV$^2 < Q^2 < 70$ GeV$^2$, with a mean
$Q^2$ of 10.7 GeV$^2$ [7].
They used a polarised muon beam, which was generated by the decay of pions
and kaons produced by primary protons
taken from the CERN S$p {\overline p}$S.
Muon beam energies of
100 GeV, 120 GeV and 200 GeV were used in the experiment.
The target consisted of two sections polarised in opposite directions.
Each was
exposed simultaneously to the beam to reduce systematic errors.
The beam polarisation was calculated as $(82 \pm 6)$\% and the
target polarisation was typically 75-80\%.
The data on $A(x,Q^2)$ was combined with measurements
of the unpolarised structure functions to determine $g_1(x)$.

EMC found no significant $Q^2$ dependence in their own $x$ bins nor when
they compared with the previous experiments from SLAC,
which took data
in the range x between 0.18 and 0.70 and $Q^2$ between
3.5 GeV$^2$ and 10 GeV$^2$.
Hence, they combined the two pieces of data assuming no $Q^2$ variation
between them.
The $g_1^p$ data from EMC and SLAC is shown in Fig. 2.
A smooth Regge extrapolation [28]
($g_1 \sim x^{-0.12}$ as $x \rightarrow 0$) was made to small x.

Thus, EMC found
$$
\int_0^1 dx g_1(x) = 0.126 \pm 0.010 (stat.) \pm 0.015 (syst.),
\eqno(19)
$$
which corresponds to the quark ``spin content"
$$
\Delta u + \Delta d + \Delta s = 0.120 \pm 0.094 (stat.) \pm 0.138 (syst.)
\eqno(20)
$$
at $<Q^2>=10.7$GeV$^2$.
This result is consistent with zero and two standard deviations from the
Ellis-Jaffe hypothesis, which says that strange quarks should not play a
significant role.
For the various flavours, one finds
$\Delta u = 0.78$, $\Delta d = -0.47$ and $\Delta s = -0.19$.
In Fig. 3
we show the convergence of the first moment
$\int_{x_m}^1 dx \ g_1(x)$ as a function of $x_m$.
The large errors on the
small x data points in the EMC $g_1^p$ data do not have a
significant impact on the integral.

This EMC spin effect (or proton ``spin crisis") has inspired a large amount
of theoretical work aimed at understanding the spin structure of the proton.
It has also
been the genesis of a new experimental program in polarised DIS.

One of the experimental priorities is to reduce the errors on the small x
data points.
Close and Roberts [29] suggested caution in adopting the Regge extrapolation
at $x \sim 0.02$
and also examined the effect of SU(3) flavour violations on the $F/D$ ratio
and the subsequent extraction of the ``spin content".
It is clearly important to have accurate small x data, although it is not
clear that
one can resolve a problem with an operator sum-rule
with data in the region $\ln {1 \over x} \gg \ln Q^2$,
where the operator product expansion analysis is not applicable.
The Spin Muon Collaboration (SMC) at CERN will take precise small x data with
a polarised proton target in 1994.
The HERMES experiment at HERA [30] will work at the same x range as EMC and
should reduce the experimental error by a factor of 10.

The other experimental priority is to measure the spin dependent neutron
structure function $g_1^n$ and test the Bjorken sum-rule.
This has been the focus of the two new pieces of data announced this year.
The SMC [21] have measured $g_1^d$ for a deuteron target in the kinematic range
$0.006 < x < 0.6$ and 1 GeV$^2 < Q^2 < 30$ GeV$^2$.
They compare this data with the EMC proton data and this enables them
(modulo small nuclear corrections)
to extract information on the neutron structure function $g_1^n(x)$.
SMC found
$$
\int_0^1 dx (g_{1,EMC}^p - g_{1,SMC}^n) = 0.20 \pm 0.05 (stat.)
\pm 0.04 (syst.),
\eqno(21)
$$
which agrees well with the theoretical Bjorken sum-rule prediction
$0.191 \pm 0.002$, to $O(\alpha_s)$ at $Q^2 = 4.6$ GeV$^2$.

On the other hand, the E-142 experiment at SLAC [22] used a $^3 He$ target
to measure $g_1^n$.
Their kinematic range was $0.03 < x < 0.6$ at an average $Q^2 = 2$GeV$^2$,
which is lower than EMC and SMC.
The polarised $^3 He$ target is believed to act as a good model of the
polarised neutron.
E-142 used the same analysis
as we outlined in section 2 to test the Ellis-Jaffe sum-rule. They used the
perturbative Wilson coefficients evaluated to order $\alpha_s$ and
obtained
$$
(\Delta u + \Delta d + \Delta s)_{SLAC} = 0.57 \pm 0.11,
\eqno(22)
$$
corresponding to $\Delta s= -0.01 \pm 0.06$ at this $Q^2$,
where $\alpha_s = 0.385 \pm 0.1$.
This number is in agreement with the Ellis-Jaffe prediction!
However,
when E-142 combined their neutron data with the EMC proton data (evolved to
$Q^2 = 2$GeV$^2$) they found
$$
\int dx (g_{1, \ EMC}^p - g_{1, \ SLAC}^n) = 0.146 \pm 0.021.
\eqno(23)
$$
This result
is a two standard deviation discrepancy from the Bjorken
sum-rule truncated to $O(\alpha_s)$ -- which evaluates to $0.183 \pm 0.007$
at $Q^2=2$ GeV$^2$.

One might question the size of the error bar on equs.(22) and (23) since
E-142 work only at $O(\alpha_s)$ and assume that the Bjorken sum-rule is exact
at an
intermediate stage of their analysis.
(They assume isospin and use $\Delta u - \Delta d = g_A^3$ for the neutron
in order to test the Ellis-Jaffe sum-rule.
Hence, the errors on equs.(22) and (23) are not totally consistent.)

The apparent discrepancy between the two experiments done
at different $Q^2$
suggests that
we have to be careful about the $Q^2$ dependence of $g_1$.
The higher order radiative
corrections to the Bjorken sum-rule (to order $\alpha_s^3$)
which we quoted in equ.(17) were calculated by Larin, Gorishny and Vermaseren
[26]. They are important at $Q^2=2$GeV$^2$,
where
the $O(\alpha_s^2)$ and $O(\alpha_s^3)$ corrections sum
to give a nett correction, which is 60\% of the size of
the $O(\alpha_s)$ correction to the Bjorken sum-rule and
each of these corrections carry the same negative sign.
The theoretical prediction for the Bjorken sum-rule is $0.164 \pm 0.009$ at
$O(\alpha_s^3)$
with $Q^2=2$GeV$^2$. This is clearly in agreement with equ.(23), within
the experimental errors.

Ellis and Karliner [31] (see also Close and Roberts [32]) have pointed
out that there may be
higher twist effects present at the momentum transfer of the SLAC experiment
and ask: ``has scaling truely set in?"
Nuclear corrections in $^3 He$ [33] may also be important in understanding the
SLAC data.
On the other hand,
it is interesting to note that the EMC and SMC data are
taken above the charm threshold for much of their x range, whereas the E-142
experiment is consistently below charm threshold.
Charm quarks turn on rapidly as one goes through threshold in the unpolarised
structure function [34].
It is tempting to think that the same is true with $g_1$ [35].
Brodsky has emphasised that there may be an intrinsic charm component in the
proton at finite x
and that this charm would be strongly polarised [36].
It would be interesting to measure $J/\Psi$ production in polarised DIS
following NMC [37]
- only as a function of the target polarisation.

Our theoretical discussion to this point has focussed on $g_1$ in the naive
parton model,
which is pre-QCD.
Just as the discovery that the quark partons only carry 50\% of the proton's
momentum led to the discovery of the
gluon distribution
so the EMC spin effect has led to a new understanding of the role of spin in
QCD.
There has been a considerable theoretical effort in the last five years trying
to understand the violation of the Ellis-Jaffe sum-rule in the EMC data.
Kodaira [25] had previously pointed to the role of the axial anomaly [38, 39]
in polarised
deep inelastic scattering.
After the announcement of the EMC data there was a
considerable theoretical debate
[40-54] aimed at understanding the exact
role of the anomaly and how it
contributes to the first moment of $g_1^p$.
More recently [55, 56],
it has been shown that the anomaly is relevant to the whole spin dependent
structure function.
The EMC spin effect should be thought of as an all moment problem!

Unfortunately, much of the discussion of the anomaly in the literature
has been rather formal. In the next section we shall endeavour to
explain the idea behind its role in polarised deep-inelastic scattering
in terms of simple quantum mechanics.

\section {Theory: the structure of the proton}

One of the main theoretical problems when analysing deep inelastic scattering
and proton structure
is to find a sensible way to break the proton up into quark and gluonic
components.
The spirit of
QCD motivated models of the proton is that we consider the three valence
quarks in some confining background field,
which is associated with condensates in the vacuum.
These potential models consist of a marriage of classical field theory
and relativistic quantum mechanics, in the spirit of Dirac's hole theory.

To see how they compare with quantum field theory we follow ref. [57]
(see also [58]) and consider the quarks in an
external gluonic field.
The traditional picture uses a Fock state expansion of the proton,
where vacuum effects
with momentum greater than some large ultraviolet
cut-off $\lambda$ are ignored.
However, this recipe leads to an
energy-momentum tensor, $\theta_{\mu \nu}$, which is not conserved
at $O(\alpha_s)$.
There is a nett flow of energy from the world of finite momenta
over the
$\lambda$ cut-off in momentum space and into the vacuum [57].
The analogous situation in classical physics is a system in thermodynamic
equilibrium with a large heat bath.
There can be a nett flow of energy between the system and the heat bath with
the system at equilibrium.
Whilst the {\it bare} energy in this process is not conserved one can write
down a {\it free} energy, which is conserved.
In quantum field theory, the
physical $\theta_{\mu \nu}$ is obtained by adding in a reciprocal
{\it regulator} term, which
represents a flux of energy and momentum from the vacuum back to the world of
finite momenta.
This physical $\theta_{\mu \nu}$ plays the role of the free energy in
thermodynamics.
However, there is a trade off: we obtain an effective conserved energy
momentum
tensor but this $\theta_{\mu \nu}$ is no longer traceless for massless quarks.
One finds
$$
\theta^{\mu}_{\mu} = (1+\gamma_m) \sum_q m_q {\overline q} q +
{\beta(\alpha_s) \over 4 \alpha_s} G_{\mu \nu} G^{\mu \nu},
\eqno(24)
$$
where $m_q$ is the running quark mass and $\gamma_m$ is the mass anomalous
dimension, $G_{\mu \nu}$ is the gluon field tensor and $\beta (\alpha_s)$
is the QCD beta function.
The gluonic term is the trace anomaly, which is very important in QCD.

Classical physics is scale invariant: there is no running coupling constant.
This scale symmetry is manifest
via Noether's theorem in terms of the partially conserved dilation current
$\partial^{\mu} s_{\mu} = \theta^{\mu}_{\mu}$.
The trace anomaly signifies a scale violation in quantum field theory.
It is intimately related to the
running of $\alpha_s$ and thereby to the scale dependence of
the parton distributions, which are measured in deep inelastic scattering.
Furthermore, the proton mass
is determined by the trace anomaly [59, 46].
The matrix element of $\theta_{\mu \nu}$ in the proton is
$$
<p| \theta_{\mu \nu} |p> = 2 p_{\mu} p_{\nu}.
\eqno(25)
$$
Taking the trace of this equation we find
$$
<p| \theta_{\mu}^{\mu} |p> = 2 m_p^2.
\eqno(26)
$$

Clearly, gluons are important here: in a free quark model with no
glue the ``proton" would be three
massless, unconfined quarks with total mass zero.
In a semi-classical quark model (say the MIT bag model) this gluonic
component appears as the infinite confining potential well in which the
quarks live.
When we say that the three valence quarks are at large $x$ in DIS
(viz. that they carry a lot of the proton's momentum) we should
remember that this proton mass or momentum is generated via the trace
anomaly.
The valence quarks are observed to carry 39\% of the proton's momentum
at $Q^2=15$GeV$^2$ [18].
This means that in the limit of massless quarks
$$
2 m_p^2 \int_0^1 dx \ x(u_v + d_v)(x)
= 0.39 <p,S|{\beta(\alpha_s) \over 4 \alpha_s} G_{\mu \nu} G^{\mu \nu}|p,S>
\eqno(27)
$$
In this sense the structure function $F_2$ measures
the distribution of the trace anomaly
over x.

The flux of quanta between the quarks with finite momenta and the vacuum
also induces a second effect (the axial anomaly),
which is measured in polarised deep inelastic scattering.
In quantum field theory the conserved axial vector current
is not invariant under gauge transformations of the external field!
The physical (gauge invariant) current differs from the conserved (free)
current by a regulator term,
which corresponds to a leakage of spin (or chirality) over the cut-off
$\lambda$ in momentum space
to the background gluonic field [57, 59].
This is the axial anomaly. It induces a two loop anomalous dimension in
the physical axial vector current, which means that this current does
not measure
the quark spin content of the proton.

In the naive parton model we can write down the spin operators for quarks
in their rest frame [60]
$$
S_k = \int d^3 x \ ({\overline q} \gamma_k \gamma_5 q), \ \ \ \  k=1,2,3.
\eqno(28)
$$
The $S_{k}$ are time independent for massless quarks and satisfy the
algebra of SU(2) spin.
In QCD, the anomalous dimension means that the flavour singlet axial current
at a scale $Q^2$ is
$$
j_{\mu, 5}^0(Q^2) = j_{\mu, 5}^0(Q_0^2) \ Z_5(Q_0^2 \rightarrow Q^2),
\eqno(29)
$$
where $Z_5$ is the renormalisation group factor.
Hence, even if the algebra of SU(2) spin is satisfied at some
scale, $Q_0^2$, at $Q^2$ we find:
$$
[S_i^{QCD}, S_j^{QCD} ](Q^2) = i \epsilon_{ijk} S_k^{QCD}(Q_0^2) \
Z_5(Q_0^2 \rightarrow Q^2).
\eqno(30)
$$
Thus the $S_k$ cannot satisfy SU(2) in QCD!
We now discuss how the axial anomaly appears in $g_1$.

\section {\bf $g_1$ in QCD}

Deep inelastic scattering provides us with a microscope, which we use
to probe the QCD structure of the proton.
The resolution is determined by the momentum transfer squared $Q^2$ carried
by the hard photon.
In order to analyse DIS in QCD we use the operator product expansion
and QCD evolution.
The flavour singlet part of $g_1$ receives contributions
from both quark and gluon partons, viz.
$$
g_{1}(x, Q^{2}) |_S = \\
{1 \over 3} \sqrt {2 \over 3} \int_x^1 {dz \over z}
\Biggl[ \Delta q_0 (z, Q^2) C^{q} ( {x \over z}, \alpha_s)
+ {1 \over \sqrt{6} } \Delta g(z, Q^2) C^g ({x \over z}, \alpha_s) \Biggr]
\eqno(31)
$$
The C-even, spin dependent quark $\Delta q (x, Q^2)$
and gluon $\Delta g(x, Q^2)$ distributions contain all the information
about the target.
They have the interpretation that they determine the probability for
finding a quark or gluon respectively, which carries a fraction $x$ of the
plus component of the target's momentum ($p_+$).
The coefficients $C(x,\alpha_s)$ are target independent and describe
the interaction of the hard photon $\gamma^{*}$ with each of these quark and
gluon partons.
We now explain the physics behind this formula.

The problem of seperating the proton into quark and gluonic components
has two ingredients.
We first need to make sure that our quark and gluon distributions
have no infra-red divergences.
Secondly, we have to renormalise the proton wavefunction
in a manner consistent with the symmetries of QCD at strong coupling.

If we work in perturbation theory, then the first of these problems
is solved by the factorisation theorem.
Consider DIS from a quark or gluon target to some order in $\alpha_s$.
This cross section has logarithmic mass singularities, which arise
from parton branching in the collinear direction (ie. at vanishing $k_T^2$).
The mass singularities are an infra-red effect.
They determine the logarithmic factors (anomalous dimensions) which govern
the QCD evolution of the parton distributions.
The great achievement of the parton idea was the proof by Amati, Petronzio,
Veneziano and others [61] (see also the earlier work [62]) that these mass
singularities
factorise to all orders in perturbation theory.
We can write the cross section for DIS from a quark or gluon target
as the convolution
of a term which contains all the mass singularities
with a coefficient term which is ``hard", by which we mean that it is
infra-red safe.
Since there are no free quarks and gluons in nature,
the confinement mechanism must somehow act to smooth out the infra-red
mass singularities.
The factorisation theorem allows us to write the structure function in
the form of equ.(31),
where we identify the infra-red term with the parton distributions.

Having taken care of the mass-singularities we then have to renormalise
the proton wavefunction
or, equivalently, the parton distributions in equ.(31).
At this point it is helpful to consider
the role of the uncertainty principle in DIS.
This is important when the hard photon interacts with a gluonic component
in the proton,
where this interaction takes place entirely
within one Compton wavelength of the photon.
The photon can resolve only that it has made a pointlike interaction
with an electrically charged parton within this distance.
It cannot resolve
whether it has interacted with a quark or with a gluonic component in the
target.

A natural signal for the interaction with a gluon
is the perturbative photon gluon fusion
process, which generates two quark jets with large
$k_T^2 \sim Q^2$ according to one's favourite jet algorithm [64].
When we calculate photon gluon fusion we start by treating the gluon as
the target.
Away from perturbation theory it is not clear (see below)
that this approximation
extends to when the hard photon makes a local interaction with the classical
background glue,
which is associated with the confining potential in QCD inspired models of
the proton.
This second process may have a different jet signature from the perturbative
photon gluon fusion.

The uncertainty principle tells us that there is no unique separation
of the local $\gamma^*$(QCD) interaction with the quarks or with
the background field in which they sit.
In unpolarised DIS the uncertainty principle is just the renormalisation
scheme dependence
of parton distributions.
For example, we can renormalise the proton wavefunction using either the
$MS$ or $\overline{MS}$ schemes.
Depending on how we have defined a ``gluon", the local $\gamma^* g$
interaction is shuffled between
the coefficient $C^g(x, \alpha_s)$ (which measures the interaction of the
photon with a gluon parton) and a local coupling (via $C^q(x, \alpha_s)$)
to the quark distribution $(q+ \overline{q})(x, Q^2)$ or proton wavefunction.
(The parton model assumes
that we can choose the coefficient to be equal to the hard photon-parton
cross-section in the full theory -- when we go beyond perturbation theory.)
The structure functions $F_1$ and $g_1$ are measured with transverse photons,
whereas
$F_2$ receives contributions from both transverse and longitudinal photons.
The polarisation direction of the photon sets the uncertainty in the local
interaction.
If the photon is longitudinally polarised the polarisation is orthogonal
to the transverse
plane and there is no uncertainty:
the longitudinal coefficients $C^q_L$ and $C^g_L$ are scheme independent
(see eg. the text by Roberts [65]).

At this point it is worthwhile to make a few remarks about the parton
model in QCD, where
the coefficients are taken to be the hard
photon-parton cross-sections.
In perturbation theory the mass singularities factorise universally
to all high-energy, inclusive
hadron scattering processes [62].
This result suggests that we may use
the parton distributions measured in deep inelastic scattering
experiments
to predict cross-sections for other high-energy processes, which are given as
the convolution
of these parton distributions with the cross-sections for parton sub-processes
leading to the requisite final state.
This parton model hypothesis is exact in perturbation theory so any violation
of it
would be a signal of non-perturbative physics -- and perhaps provide a window
on the background field (or QCD vacuum) which leads to confinement.
The parton model
has been tested to $O(\alpha_s^3)$ precision in unpolarised
$p {\overline p}$ scattering at CDF, where there is excellent agreement
between theory and data [63].
However, as we explain below, the parton separation between hard and soft
is less reliable in spin dependent processes.

In polarised DIS some of the uncertainty is frozen out by the gauge symmetry
of QCD.
The parton distributions (or proton wavefunction) in $g_1$ are defined
by the operator product expansion so that their odd moments project out
the target matrix elements of the {\it renormalised}, spin-odd, composite
operators
$$
2m_p S_+ (p_+)^{2n} \int^1_0 dx \ x^{2n} \Delta q_{k} (x, Q^2) =
<p,S | [ {\overline q}(0) \gamma_+ \gamma_5 (i D_+)^{2n} {\lambda^k \over 2}
q(0) ]_{Q^2}^{GI} |p,S >_c
\eqno(32)
$$
$$
2m_p S_+ (p_+)^{2n} \int^1_0 dx \ x^{2n} \Delta g (x, Q^2) =
<p,S | [ {\bf \rm Tr} \ G_{+ \alpha}(0) (iD_+)^{2n-1}
{\tilde G}^{\alpha}_{\ +}(0) ]_{Q^2}^{GI} |p,S >_c
(n \geq 1)
\eqno(33)
$$
Here $D_{\mu} = \partial_{\mu} + ig A_{\mu} $ is the gauge covariant
derivative in QCD.
The association of $\Delta q_k(x, Q^2)$ with quarks and $\Delta g(x, Q^2)$
with gluons follows
when we evaluate the target matrix elements in equs.(32) and (33)
in the light-cone
gauge, where $D_+ \rightarrow \partial_+$.
(The superscript on the operators emphasises that they are quoted with
respect to some gauge invariant regularisation scheme,
and the subscript $Q^2$ indicates the subtraction point.)

The parton distributions are renormalised quantities which should always
be quoted with respect to some ultraviolet regularisation.
As we explained in section 5,
the choice of regularisation determines the symmetry properties of the
theory.
The clash of symmetries between different regularisations is what is meant
by an anomaly.
(For a complete discussion of regulator theory we refer to the text of
Zinn-Justin [66] and the lectures of Guichon [67].)
The physical axial vector current does not measure a quark spin content
because of its anomalous dimension.
In classical physics the axial vector current is gauge invariant: the quark
field transforms as
$$
q(x) \rightarrow U(x) q(x)
$$
and
$$
{\overline {q}}(x) \gamma_{\mu} \gamma_{5} \rightarrow
  {\overline {q}}(x) \gamma_{\mu} \gamma_{5} U^{\dagger}(x)
\eqno(34)
$$
under a given gauge transformation $U$.
On the other hand, in quantum field theory the axial vector current
operator is not just
${\overline q}(0) \gamma_{\mu} \gamma_{5}$ multiplied by $q(0)$.
It is a composite operator which has to be
renormalised and there are extra
divergences which are intrinsic to the operator itself.
The conserved axial vector current, which does measure spin, is the
symmetry current [68]
$$
j_{5 \ \mu}^S = j_{5 \ \mu}^{GI} - k_{\mu},
\eqno(35) $$
Here
$$
k_{\mu} = {\alpha_s \over 2 \pi}
\epsilon_{\mu \lambda \alpha \beta} {\bf Tr} \ A^{\lambda}
(G^{\alpha \beta}-{2 \over 3} A^{\alpha} A^{\beta})
\eqno(36)
$$
is a gauge dependent gluonic current, which measures the leakage of spin
to the background field.
The gauge dependent symmetry current $j^S_{5 \mu}$ is {\it the}
axial-vector current
in a world where chirality is
an exact symmetry and where gauge invariance is not.
The anomaly in the axial vector current is important in QED, where it
predicts the decay rate for $\pi^0 \rightarrow 2 \gamma$ [38, 39].
This decay would be strongly suppressed without the anomaly.
The axial anomaly is also relevant to our understanding of the U(1)
problem in QCD [69].

It is important to realise that the anomaly is not restricted to an effect
in the axial-vector current.
An anomaly will manifest itself
wherever the quantum numbers of the relevant operator allow it.
For example,
any renormalisation calculation involves a subtraction scale - and this
breaks the scale invariance of the classical theory.
The only operators which are scale invariant are those which are not
renormalised (ie. conserved currents and the other ``soft" operators
- the divergences of which just involve mass terms).
The QCD evolution of the parton densities with increasing $Q^2$ and
asymptotic freedom
are intimately related to the scale anomaly.
It turns out that one cannot renormalise each of the $C=+1$ axial tensor
operators
in a gauge invariant way so that they describe spin at the same time.
The operator which does measure spin
differs from the corresponding gauge invariant operator by a multiple
of a gauge-dependent, gluonic counterterm $k_{\mu \mu_1 ... \mu_{2n}}$,
viz.
$$
\biggl[
{\overline q}(0) \gamma_{\mu} \gamma_5 D_{\mu_1} ... D_{\mu_{2n}} q(0)
\biggr]^S_{Q^2} =
\biggl[
{\overline q}(0) \gamma_{\mu} \gamma_5 D_{\mu_1} ... D_{\mu_{2n}} q(0)
\biggr]^{GI}_{Q^2}
+ \lambda_{S, n} \biggl[ k_{\mu \mu_1 ... \mu_{2n} } \biggr]_{Q^2}
\eqno(37)
$$
where the coefficients $\lambda_{S, n}$ and anomalous currents were
discussed in ref. [55].

The operator product expansion analysis of deep inelastic scattering
involves only gauge-invariant {\it operators},
whence it follows that the spin dependent
quark distribution in DIS does not measure the quark spin content of the
proton.
(Gauge invariance wins out over chirality!)
Instead, it includes a local interaction between the hard photon and
the background field.
One can say that {\it the gauge symmetry screens the spin of the quarks.}

Since the only operator which contributes to the first moment of $g_1$ is
the gauge-invariant axial-vector current
and this operator is associated with the quark distribution $\Delta q(x,Q^2)$,
it follows that
$$
\int_0^1 dx \Delta g(x,Q^2) \ \int_0^1 dz C^g (z,\alpha_s) = 0
\eqno(38)
$$
In general, $\Delta g(x,Q^2)$ is target dependent, whence one concludes that
the first
moment of the spin dependent gluonic coefficient must vanish to all orders
in perturbation theory at the twist-two level [55] (see also [48]).

At this point it is important to understand the physical implications of
this discussion.
In the parton model we would like to set the gluonic coefficient equal
to the hard cross-section for photon-gluon fusion.
This means including the local $\gamma^* g$ interaction, which is associated
with the anomaly,
into $C^g(x, \alpha_s)$, whereby the first moment of $g_1$ would measure the
spin of the quarks minus
a contribution from the polarised gluon distribution $\Delta g(x,Q^2)$,
viz. $-{\alpha_s \over 2\pi} N_f \Delta g$ [41-44].
However, this corresponds to a parton model, where the quark distribution
is defined
with respect to the S-prescription axial tensor operators.
(That is, it lives in the world where chirality is an exact symmetry at the
expense of gauge-invariance.)
It is well known from the work of Jaffe and Manohar [46] that the forward
matrix elements
of $k_{\mu}$ (and by extension the higher spin operators,
$k_{\mu \mu_1 ... \mu_{2n} }$) are not gauge-invariant beyond perturbation
theory.
In otherwords, the parton model corresponds to a regularisation which breaks
the gauge symmetry
at strong coupling when we consider spin-dependent processes.

Nachtmann and collaborators [71] have pointed out that the parton model
hypothesis
may also breakdown
when one considers spin in the Drell-Yan process
due to non-perturbative aspects of the background field inside hadrons.
They point to evidence of this effect in the NA-10 data [72].
The parton model idea that we can make a clear separation of
high-energy, hadronic scattering
processes into hard and soft
fails when we consider spin dependent processes.

\section {Phenomenology of the anomaly in $g_1$}

We now illustrate this discussion by comparing $g_1$ with the other structure
functions, which are measured in DIS experiments.
The axial anomaly is relevant only to $g_1$.
It is not present in the unpolarised quark distributions,
which are described in OPE language by the operators
$ {\overline q}(0) \gamma_+ (iD_+)^{n} q(0)$.
Nor is it relevant to the C-odd spin dependent structure function
$g_3$, which is the polarised version of $F_3$.
(In the naive parton model $g_3$ measures the spin dependent valence
distribution
$(q-{\overline q})^{\uparrow}(x) - (q-{\overline q})^{\downarrow}(x)$.)
Since $g_3$ is odd under charge conjugation,
it can have no anomalous gluonic contribution.
(This result follows from the Furry theorem.)
This means that it does make sense to talk about $F_1$, $F_3$ and $g_3$
in terms of quarks with explicit spin degrees of freedom - the clash of
symmetry between gauge invariance and spin does not manifest itself in
these structure functions.

In order to deal with $g_1$ we should modify the parton spin
identification in equ. (12) by writing
the gauge invariant distribution as [56, 73]
$$
q^{\uparrow}_{GI}(x, Q^2) = ( q^{\uparrow}_S + {1 \over 4} \kappa ) (x, Q^2)
\eqno(39a)
$$
$$
q^{\downarrow}_{GI}(x, Q^2) = ( q^{\downarrow}_S - {1 \over 4} \kappa ) (x,
Q^2)
\eqno(39b)
$$
for both quarks and anti-quarks.
Here $\kappa (x, Q^2)$ denotes the anomaly and the subscript $S$ denotes
the gauge-dependent spin distribution.
The $\kappa$ distribution
appears only in the treatment of $g_1$ in deep inelastic scattering.
Since the anomaly is independent of quark flavour the same $\kappa$
distribution is relevant to each of $u, d, s, c, ..$.
It varies only according
to the ``spin" and not the charge
or flavour.
Because $\kappa$ is flavour independent it will induce some OZI violation
wherever it plays a role.
This is the likely source of the EMC spin effect.

If we substitute equs.(39) into the OPE expression for $g_1$ (viz. equ.(31))
 it is easy to see that both the spin and anomalous components of
$\Delta q (x, Q^2)$ couple to the
hard photon in exactly the same way as one expects of a quark
(via $C^q(x,\alpha_s)$).
Physically, this means that there is a
new local interaction between the hard photon and a gluonic component in
the proton, which must be included in the
parton model [54, 55]
This is despite the fact that the glue does not carry electric charge !

We now discuss the renormalisation group properties of the anomaly
in $g_1(x, Q^2)$.
At $O( \alpha_s )$ the measured parton distributions evolve with
increasing $Q^2$
according to the GLDAP equations [74]
$$
{d \over dt} \Delta q_0 (x, t) = {\alpha_s \over 2\pi }
\int_x^1 {dz \over z} \Biggl[ \Delta q_0 (z, t) P^q_{qq} ({x \over z}) +
2 N_f \Delta g (z, t) P^q_{gg} ({x \over z}) \Biggr]
$$
$$
{d \over dt} \Delta g (x, t) = {\alpha_s \over 2\pi }
\int_x^1 {dz \over z} \Biggl[ \Delta q_0 (z, t) P^g_{qq} ({x \over z}) +
\Delta g (z, t) P^g_{gg} ({x \over z}) \Biggr]
\eqno(40)
$$
where $t = \ln Q^2$.
The splitting functions have a probability interpretation in the light-cone
gauge so that eg.
$P^q_{gg}(x)$ is the probability
for a gluon to branch into a quark-antiquark pair so that the quark
(or antiquark) carries a fraction $x$ of the gluon's momentum in the
collinear direction.
The splitting function is the coefficient of $\ln {Q^2 \over \mu^2}$
in the cross section for DIS from a quark or gluon at $O(\alpha_s)$.

In section 6 we explained that the anomaly $\kappa$ appears as a local
interaction of the hard photon with a gluon, where the interaction takes
place within one Compton wavelength of the photon.
In perturbation theory, this corresponds to $k_T^2 \sim Q^2$ and there is
no logarithm at $O(\alpha_s)$.
This means that we can disentangle an evolution equation for $\kappa$ [56]
from the GLDAP equations:
$$
{d \over dt} \kappa (x,t) = 0.
\eqno(41a)
$$
All of the QCD evolution of $\Delta q(x, Q^2)$ at $O(\alpha_s)$
is carried
by its spin component.
If we substitute equs.(39) into equs.(40) we can re-write the GLDAP
equations as
$$
{d \over dt} \Delta g(x, t) =
{\alpha_s \over 2 \pi} \int_x^1 {dz \over z}
\Biggl[ ( \Delta q_S (z, t) + \kappa (z, t) )
P^g_{qq} ({x \over z}) + \Delta g(z, t) P^g_{gg} ({x \over z}) \Biggr]
\eqno(41b)
$$
$$
{d \over dt} \Delta q_S(x, t) = {\alpha_s \over 2 \pi}
\int_x^1 {dz \over z} \Biggl[ (\Delta q_S (z, t) + \kappa(z, t) ) P^q_{qq} ({x
\over z}) +
2 N_f \Delta g(z, t) P^q_{gg} ({x \over z}) \Biggr]
\eqno(41c)
$$
The anomaly $\kappa (x, Q^2)$ scales at $O(\alpha_s)$ and has a different
evolution equation from the partonic gluon distribution $\Delta g(x,Q^2)$.
Clearly, the anomaly is an entirely different gluon effect in $g_1$
than $\Delta g(x, Q^2)$.

We can now write down the properties of the anomaly in $g_1$.
\begin{itemize}
\item
The background field (or confining potential) in QCD is important if we
wish to understand the shape of the
structure functions in deep inelastic scattering.
The trace and axial anomalies describe a flux of quanta between the quarks
with finite momenta and the vacuum.
The trace anomaly determines the proton mass in terms of the confining
potential.
The axial anomaly means that the spin of the quarks is screened by the
background field in which they live.

\item
In general, the spin and anomalous components of $\Delta q(x,Q^2)$ are not
separately gauge invariant.
Rather than try to separate $\Delta q(x,Q^2)$ into meaningful quark and
gluonic components
we take the view that one should keep with the gauge invariant distributions
in equ.(31), while bearing
in mind that $\Delta q(x,Q^2)$ contains some OZI violation associated
with the axial anomaly.

\item
The anomaly in $\Delta q(x,Q^2)$ scales at $O(\alpha_s)$ and couples to
the hard photon in
exactly the same way as a quark.
Hence the anomaly makes a scaling contribution to $g_1$.
\end{itemize}

It is clearly an important experimental challenge to map out the x dependence
of the OZI violation in $g_1$.
In section 5 we explained how the unpolarised structure function can be
viewed as the distribution of the trace anomaly over x: the trace anomaly
is relevant to large x!
Formally, it is clear that the axial anomaly is also important at large x
since
it is intrinsic to each of the higher moments of $\Delta q(x,Q^2)$.
Since the anomaly is not present in the spin-dependent valence distribution,
$g_1$ and the valence part of $g_1$ probe different aspects of what we call
the constituent quark.
The interesting
phenomenology of a large x anomaly is a possible violation of Zweig's rule
in the large x bins of the polarised DIS experiment.
Of course, the size of this OZI violation at large x is a problem for
experiment.
The point is that there is no good theoretical reason to expect that the
anomaly is constrained to small x.
A large x anomaly
would mean that the spin of the three valence quarks is screened by the
confining potential in $g_1$.
It would be seen as a violation of Zweig's rule in the large x bins of the
polarised DIS experiment.
This would be in contrast with the ordinary gluon
distribution $\Delta g(x,Q^2)$,
which is relevant to $g_1$ only as small x ($x \leq 0.03$) [47, 49, 50].

To see that ultraviolet regularisation effects (anomalies) can be present
at large x
it is also helpful to consider the light-cone correlation function in polarised
DIS.
At the semi-classical level we can write $\Delta q(x, Q^2)$ as the Fourier
transform
$$
\Delta q(x,Q^2) = {1 \over 2\pi} \int dz_- \cos (x p_+ z_-)
<p, S| \overline{q}(z_-) \gamma_+ \gamma_5 q(0) |p, S>,
\eqno(42)
$$
where this object requires careful renormalisation in the full quantum field
theory [75].
By inverting the Fourier transform for typical valence and sea distributions
one finds [75] that
the correlation function for the valence distribution is peaked about $z=0$,
whereas large correlation lengths ($z_- > 1$fm) are associated only with small
x.
Small $z_-$ contributes to the full range of x.

At this stage we consider Schwinger's derivation of the anomaly [76]
using point
splitting regularisation.
The physical axial vector current is obtained as the $z \rightarrow 0$ limit
of
$$
{\overline q}(z) \gamma_+ \gamma_5 e^{ie \int A_{\mu}dy^{\mu}} q(0)
\eqno(43)
$$
acting in an external field.
Modulo the path ordered exponential this is the same object that appears in
the correlation function equ.(42).
In the Schwinger derivation $z$ acts as the ultraviolet regulator and the
anomaly appears as a term $\sim {1\over z}$ in the expansion of equ.(43) about
$z=0$.
The anomaly is associated with small correlation lengths and therefore has the
potential to contribute at large x.

\section {Modelling the axial anomaly in $g_1$}

Since the anomaly is not present in the valence (C-odd) part of $g_1$
it would be most helpful
to obtain data on the C-odd spin dependent proton structure function.
A large $x$ anomaly
would show up as a finite difference between $g_1$ and its valence part
at large $x$ [55, 73].
This difference should be associated with a second gluon distribution
in polarised DIS,
which is a clear window on the confining potential in the proton.
If the anomaly is purely a small x effect it would not be possible to
isolate it from the sea and the ordinary gluon distribution $\Delta g(x, Q^2)$,
which dominate the small x
data.
(We shall come back to the ordinary gluon distribution at the end of this
section.)

Unfortunately, the cross section for DIS with a neutrino beam and proton
target is very small - enough to
make direct measurements of $g_3$ impracticable at the present time.
However,
in the naive parton model it is possible to extract the C-odd valence
distribution from
$g_1$ measurements by detecting fast pions from among the final state hadrons
[77, 78].
Whilst this analysis needs to be updated in the light of the anomaly
it is clear that semi-inclusive measurements provide the best chance
to measure the spin dependent
valence distribution at the present time.
This experiment is planned by the HERMES collaboration at HERA [30] --
preliminary data is also available from SMC, albeit with large errors.

Important information about the $x$ dependence of the axial anomaly
in polarised deep inelastic scattering also comes from
measurements of $g_1^n (x, Q^2)$.
The axial anomaly occurs only in the flavour singlet part of $g_1$ and
therefore it will be present equally in
$g_1^p$ and $g_1^n$ as a function of $x$.
If the anomaly acts to screen the quark spin at large $x$
in $g_1^p$
it follows that the same should be true in $g_1^n$.
The combination appearing in the Bjorken sum-rule $(g_1^p-g_1^n)$
has no flavour singlet component and is anomaly free.
On the other hand, the flavour singlet component is enhanced in the
deuteron structure function
$g_1^d ={1\over2} (g_1^p + g_1^n)$,
which has no isotriplet piece $\Delta q_3 (x, Q^2)$.
Thus the deuteron structure function $g_1^d$ is an ideal place to test
model predictions about how the anomaly should contribute in the nucleon
structure function $g_1 (x,Q^2)$.

The usual quark model calculations, which do not
include the anomaly,
suggest that $g_1^n$ will change sign and become small and positive
at large $x$ [79-82].
To the extent that these models do not include any OZI violation,
a large $x$ anomaly would tend to render $g_1^n$ negative at large $x$.
As a specific example we consider
the quark model calculation of the structure functions
which was developed by the Adelaide group [80-82] following earlier
work by Jaffe, Ross and others [83].
These calculations use the MIT bag model, a popular, relativistic model
of hadron structure, whch reproduces low energy hadronic properties
such as charge radii, magnetic moments and the flavour non-singlet axial
charges.
This same model also provides a good description of
unpolarised structure function data, notably $x F_3(x)$ and the
$d/u$ ratio.
For polarised deep inelastic scattering the model also respects the
Bjorken sum-rule and, because it has no OZI violation,
the Ellis-Jaffe sum-rule.
(The bag model has not yet been extended to satisfy the U(1) chiral Ward
identity.)
Because of this last feature it is clear that it cannot fit the EMC
data for $g_1^p$.
Nevertheless, it does seem reasonable that the bag model prediction for
$g_3^p$ should be reliable
since this structure function is free of any Zweig's rule violations
due to the anomaly.

In view of the general success of the model in unpolarised physics
and our earlier description of the role of the anomaly, which is certainly
missing
in the bag treatment, we make the working hypothesis that the discrepancy
in $g_1^p$ is entirely due to the anomaly.
By adding a purely phenomenological term to the bag model calculations
to fit the EMC data, we are then in a position to make predictions for
$g_1^n$ and $g_1^d$.
This is illustrated in Fig. 4a.
Figure 4a shows the EMC and earlier SLAC data for $x g_1^p$ together with the
naive bag model expectation
(the solid line) corresponding to a bag radius $R = 0.8$fm, which we take
from Schreiber et al. [81].
The dashed curve is the result of adding a purely phenomenological
term to fit the data [84].
We will associate this phenomenological term with the OZI violation
missing in the naive bag model.
In this case, the same flavour singlet correction should therefore be applied
to the neutron
and this is shown in Fig. 4b.
Here the solid curve is the naive bag (no OZI violation) prediction for
$x g_1^n$ (again at $Q^2 = 10$GeV$^2$ for a bag radius $R=0.8$fm) and
the dashed curve is obtained by adding the same correction that was
applied for the proton. Clearly when this correction is included
we find that the neutron spin-dependent structure
function becomes negative at large $x$.
In Fig. 4c we show the naive bag (solid) and anomaly corrected (dashed)
predictions for $x g_1^d$.
(For the present purposes we make the simple approximation that
$ g_1^d = ( g_1^p + g_1^n )/2$ ,thus ignoring corrections due to
shadowing,Fermi-motion and the D-state probability of the deuteron.
These are expected to be important at the few-percent level [85] - well below
the present experimental accuracy.)
The corrected curve is in good agreement with the recent SMC measurement
of the deuteron spin structure function $x g_1^d(x,Q^2)$ [21].

We repeat this analysis in Figs. 5a-5c for a bag radius of $R=0.6$fm.
The bag model calculation of $g_1$ is taken from ref.[79], while the
dashed curve is again the bag result supplemented by a purely
phenomenological term [86].
Again the bag model supplemented by the anomaly term
tends to favour a negative sign for $g_1^n$ at large $x$,
and there is reasonable agreement with the SMC data for $x g_1^d$.
If the anomaly is an intrinsic part of the constituent quark then one
would expect it to have the same $(1-x)^3$ behaviour,
which is given by the counting rules.
It is interesting to note that the phenomenological term in our
calculations,
which we
associate with the anomaly, roughly reproduces this shape in the middle
$x$ range where the data is most reliable.

We have stressed the difference between the anomaly and the ordinary gluon
distribution $\Delta g(x,Q^2)$.
It is known from unpolarised DIS experiments that the gluon distribution
is concentrated at small $x$. In polarised DIS
the hard photon scatters from the gluon distribution $\Delta g(x,Q^2)$
via a quark-antiquark
pair, described in $C^g (x, \alpha_s)$, viz.
$$
g_1^G (x,Q^2) = {1 \over 9}
\int_x^1 {dz \over z} \Delta g(z, Q^2) C^g ({x \over z}, \alpha_s)
\eqno(44)
$$
This dissipates the gluon's already small momentum so that
$\Delta g(x, Q^2)$ is relevant to $g_1$ only at small $x$ ($x \leq 0.03$)
[49,50].

At present we do not have a measurement of $\Delta g(x, Q^2)$
so we need to
deduce some
phenomenological parametrisation to estimate this gluon contribution in the
$g_1$ data.
The hardest possible $\Delta g(x,Q^2)$ that we can write down consistent with
the inequality
$$
|\Delta g(x) = (g^{\uparrow} - g^{\downarrow})(x)|
\leq g(x) =    (g^{\uparrow} + g^{\downarrow})(x)
\eqno(45)
$$
is
$$
\Delta g(x) = x^{\alpha} g(x)
\eqno(46)
$$
where $\alpha \geq 0$.
For definiteness, consider the experimental parametrisation [84]
$$
xg(x,Q^2) = 0.88 (1+9x) (1-x)^4
\eqno(47)
$$
at $Q^2=4$GeV$^2$.
(The EMC low $x$ data was measured at $<Q^2>=4$GeV$^2$ so we can use equ.(46)
to estimate
the gluon component in this part of the data.)
We use the $\overline{MS}$ coefficient $C^g(x, \alpha_s)$ [48, 53] and take
$\alpha =0.2$
in equ.(46), which corresponds to a large polarised gluon component
$\Delta g = 4.0$ in the proton.
(A large $\Delta g \sim 4$ has been postulated by several authors [41-44].)
The contribution of $\Delta g(x,Q^2)$
to the EMC data is shown in Fig. 6, where we plot the combination
$(g_1^{EMC data} + g_1^G )(x)$.
This plot allows us to see the gluonic contribution to the EMC data on $g_1$
at a glance.
The gluon distribution $\Delta g(x,Q^2)$ contributes to $g_1^p$ only at small
x and is clearly well within the experimental errors where data has been taken.
It makes a negligible contribution to the measured
sum rule between $x = 0.01$ and 1, where the three constituent quarks are
expected to dominate.
This result is essentially independent of the choice of the parametrisation
for $\Delta g(x)$.
(Similar calculations, using a variety of parametrisations for $\Delta g(x)$
have appeared in refs. [47,49-50,88].)
Whilst the gluon distribution $\Delta g(x,Q^2)$ does not contribute to the
first moment of
$g_1$ (equ.(38)), the sign of $\Delta g$ clearly
affects the low $x$ region ($x \leq 0.01$).
{\it With present data even the sign of $\Delta g$ is undetermined.}

In concluding this section we should record the fact that there have
been a large number of papers devoted to the calculation of corrections
to traditional models of nucleon structure that have the effect of
correcting the Ellis-Jaffe sum-rule. The nearest in spirit to the work
emphasised here is that involving perturbative corrections arising from
instantons [89, 90] -- since instantons break axial
U(1) symmetry. Additional mechanisms considered have included gluonic
and pionic corrections [91-94].
All of the latter shift the spin of the nucleon into orbital angular
momentum of nucleon constituents (pions,quarks and gluons). It is a very
important challenge to theory and experiment to understand
quantitatively the role played by such corrections in comparison with
the more fundamental explanation on which we have focussed.

\section {Elastic neutrino proton scattering }

There have been a large number of theoretical papers [46, 95-98].
suggesting that the value of $\Delta s$ extracted from the
EMC measurement of $g_1$ should be compared with
a quantity (also denoted ``$\Delta s$") which is measured in elastic
neutrino proton scattering.
This point needs some clarification.

Elastic neutrino proton scattering proceeds at leading order
through the neutral current (single $Z^0$ exchange).
In the notation of equ.(15), the $Z^0$
couples to the combination
$\Delta u - \Delta d - \Delta s + \Delta c$, which is flavour
non singlet;
the $\nu$p cross section is anomaly free.

The term $\Delta c$ is commonly omitted and this is the source
of much confusion.
An intrinsic part of the $\nu$p cross section involves the $Z^0$
coupling to the target via two gluons. This is shown in Fig. 7.
It is necessary to include charm here because all flavours of
quark can flow through the virtual loop in the vector (gluon), vector
(gluon) axial-vector ($Z^0$) subgraph.
It is not correct to ignore the charm quarks just because
of the heavy quark mass.
If the charm contribution is neglected
we cannot satisfy current conservation at each of the $Z^0$
and gluon vertices.
For a given flavour of quark, we can ensure current conservation
at the gluon vertices via the anomaly. However, we cannot then
renormalise the triangle amplitude to
ensure current conservation at the $Z^0$ vertex.
This would mean subtracting out the anomalous counterterm
which we introduced to maintain gauge invariance at the gluon
legs.
The anomaly is flavour independent.
Hence, there is no problem when we consider the full non singlet
coupling at the $Z^0$ vertex : the nasty divergences which
act to destroy gauge invariance cancel out and we have a sensible
theory.
(The apparent loss of gauge invariance at the $Z^0$
vertex due to the mass terms in the divergence equation
is restored by the Higgs mechanism.)
We must sum over the full generation of quarks
to ensure current conservation and, hence, renormalisability.
Indeed, this is the reason for demanding a top (truth) quark in
the theory - to cancel the anomaly associated with the
beautiful quark.

The value of ``$\Delta s$" which is determined in elastic neutrino
proton scattering is really $\Delta s - \Delta c$.
This combination is scale independent.
It should not be compared directly with the $\Delta s$ that is
extracted from the EMC experiment and which does have
an anomaly and is scale dependent.
The currently accepted value is
$\Delta s - \Delta c = -0.15 \pm 0.08$ [99], which
agrees within errors with the EMC value for $\Delta s= -0.19$.
However, as the experimenters themselves point out [99] (see also Close [95]),
the value of $\Delta s - \Delta c$
which is extracted from the neutrino experiments is strongly
dependent upon the dipole mass term in the axial form factor.
By varying this mass parameter within one standard deviation,
one easily finds that $\Delta s - \Delta c$ is consistent with
zero.

Although it may appear to belabour the point we feel that it is
worthwhile to make one further point concerning neutrino proton
($\nu$p) scattering.
The virtual $Z^0$ involved in this process carries some four
momentum transfer $Q^2$, so that in general the matrix
element of ${\overline q} \gamma_{\mu} \gamma_5 q$ is taken between proton
states of
different momentum. This introduces a form-factor which
as we mentioned above is usually parameterised as a dipole
(with mass parameter about 1GeV). It would be incorrect to
confuse this dependence with the scale dependence of quantities
measured in DIS, where the forward Compton amplitude involves
initial and final protons of the same momentum.
In particular, it would not be correct to suggest that at
(say) zero momentm transfer $\nu$p scattering would determine
$\Delta s$ at a scale $Q^2=0$.
The combination which is actually measured, namely $\Delta s - \Delta c$,
is scale invariant.
By itself $\Delta s$ has a scale dependence induced by the
anomaly, but if $\Delta s$ were to grow by an amount $\delta$
when changing scale from $\mu_0^2$ to $\mu^2$, then
$\Delta c$ would grow by the same amount.

The interpretation of the deep inelastic process
cross section depends on whether we are above or below the
threshold for the production of real charm.
This differs from the elastic neutrino proton scattering
where the charm quarks are purely virtual.
Whether or not we can apply the OPE is determined by the kinematics
of the physical process.
By applying the OPE we are assuming that $Q^2$ is much greater
than all the other QCD scales in the problem.
This is fine for the light quarks.
Far below charm threshold ($Q^2 \ll 10$GeV$^2$) charm quarks
will be frozen out of DIS.
Well above threshold ($Q^2 \gg 10$GeV$^2$)
they participate fully and $\Delta c^{(DIS)}$ is
given by relation to the OPE as the proton matrix element
of the axial vector current $\overline{c} \gamma_{\mu} \gamma_5 c$,
$$
2m S_{\mu} \Delta c^{(DIS)} =
<p, S| \overline{c} \gamma_{\mu} \gamma_5 c | p, S>.
\eqno(48)
$$
Near the threshold the problem is complicated - for a detailed
discussion we refer to Witten [100].
In any case, the situations for DIS and $\nu$p elastic scattering
(where the charm quarks are purely virtual) are physically
quite distinct [35].

\section{ Conclusions and outlook}

In this review we have outlined the experimental results for spin
dependent deep inelastic scattering that have generated so much
interest. The reasons for that interest in terms of the traditional
parton model were then explained. Having established this background
we were able to concentrate on the deeper theoretical issues. We made it
clear that the EMC spin effect is {\it not} a question of the fraction
of the spin of the nucleon carried by its quarks; when gauge invariance
is respected this is not a well-defined concept.

Although attention has been traditionally focussed on the violation of
the Ellis-Jaffe sum-rule -- that is, the violation of OZI in the first
moment of $g_1$ -- we have been at pains to emphasise that this throws
away a great deal of information. In particular, the conflict between
chiral symmetry and gauge invariance that lies behind the U(1) axial
anomaly occurs in every moment of $g_1$. Indeed we have argued that the
spin structure of the nucleon offers a unique window into nucleon
structure and the mechanism of confinement. At first glance this makes
our approach quite different from the approach of Shore and Veneziano [9, 45]
who argue that the OZI violation is target independent -- being related
to the scale dependence of the decay constant of the analogue of the
$\eta ^{'}$ in OZI QCD.
The approach of ref.[45] and that discussed here should be related at some
deeper level
(as both approaches stem from the conflict between gauge invariance and
chiral symmetry mentioned above).
However, there remains an open problem of how to generalise their analysis
to address higher
moments (i.e. the {\it shape} of $g_1$).

We have discussed how one could map out the $x$ dependence of
the axial anomaly in $g_1$.
If the anomaly does produce effects at large $x$ it can be isolated as a
finite difference between the structure functions $g_1$ and $g_3$
in the large $x$ bins. If it
is purely a small $x$ effect it would be lost among the sea
and gluon
distributions which dominate the data in this region (say $\leq 0.1$).
Certainly, the anomaly is an intrinsic
property of what we call the constituent quark
and there is no good theoretical reason to believe that it is confined to
small $x$.

We strongly urge that consideration be given to the challenging
experimental
problem of how to measure $g_3$.
In the interim it would be very useful to obtain more data (with reduced
errors) on the deuteron spin structure function $g_1^d$.
The HERMES experiment aims to reduce the errors on $g_1^d$ by a factor
of 10 [30].
This new deuteron data will be able to discriminate between the two
theoretical curves in Figs.4c and 5c and
will provide an important constraint on models of nucleon structure.
At the same time it is important to map out the $Q^2$-dependence of the
data, especially through the charm threshold. Finally, little attention
has so far been paid to the possible ambiguities in extracting $g_1^n$
from experiments on nuclei where the nucleons are necessarily off-shell.
It has recently been shown that this can lead to a substantial
correction in the unpolarised case [101], and mindful of Bjorken's Law (quoted
at the beginning) one must worry even more about the polarised case.

\vspace {1.0cm}

\begin{center}{\bf Acknowledgements} \end{center}

This work was supported in part by the Australian Research Council.

We wish to thank our many colleagues with whom we have had the pleasure
to discuss the spin problem
in recent years.
We are especially grateful to S. J. Brodsky, O. Nachtmann, G. M. Shore,
B. R. Webber
and members of the SMC collaboration
for helpful discussions at the concluding stages of this work.

\pagebreak

\begin{center}{\bf References}\end{center}
\vspace{1.0cm}
\begin{enumerate}
\item\label{cern}
CERN Courier, Volume 30, number 9 (December 1990) 1.
\item\label{fe69}
R. P. Feynman, Phys. Rev. Lett. 23 (1969) 1415.
\item\label{bj1}
J. D. Bjorken, Phys. Rev. 179 (1969) 1547.
\item\label{bj2}
J. D. Bjorken and E. A. Paschos, Phys. Rev. 185 (1969) 1975.
\item\label{awt1}
A. W. Thomas, Prog. Part. Nucl. Phys. (1987) 21.
\item\label{awt2}
A. W. Thomas and W. Melnitchouk,in Proceedings of the
JSPS--INS Spring School (Shimoda) (World Scientific,Singapore,1993).
\item\label{emc1}
J. Ashman et al., Phys. Lett. B206 (1988) 364, Nucl. Phys. B328 (1990) 1.
\item\label{jpg}
S. D. Bass and A. W. Thomas, J. Phys. G19 (1993) 925.
\item\label{vs90}
G. Veneziano, Okubofest lecture, CERN preprint TH-5840/90 (1990).
\item\label{el93}
J. Ellis, to appear in Proc. PANIC XIII (Perugia), World Scientific (1993).
\item\label{rw92}
R. Windmolders, Int. J. Mod. Phys. A7 (1992) 639.
\item\label{fec79}
F. E. Close, {\it An Introduction to Quarks and Partons}, Academic Press
(1979).
\item\label{ru11}
E. Rutherford, Phil. Mag. 21 (1911) 669.
\item\label{fec83}
F. E. Close, {\it The Cosmic Onion}, Heinemann (1983).
\item\label{ja90}
R. L. Jaffe, Comments Nucl. Part. Phys. 19 (1990) 239.
\item\label{gw74}
D. J. Gross and F. Wilczek, Phys. Rev. D9 (1974) 980.
\item\label{emc2}
EMC, J. J. Aubert et al., Phys. Lett. B123 (1983) 275.
\item\label{emc3}
T. Sloan, G. Smadja and R. Voss, Phys. Rep. 162 (1988) 45.
\item\label{slac1}
G. Baum et al., Phys. Rev. Lett. 51 (1983) 1135.
\item\label{emc4}
V. Hughes et al., Phys. Lett. B212 (1988) 511.
\item\label{smc}
The SMC Collaboration, B. Adeva et al., Phys. Lett. B302 (1993) 533.
\item\label{slac2}
The E-142 Collaboration, P. L. Anthony et al., Phys. Rev. Lett. 71 (1993) 959.
\item\label{bo83}
M. Bourquin et al., Z. Phys. C21 (1983) 27.
\item\label{kod79}
J. Kodaira et al., Phys. Rev. D20 (1979) 627, Nucl. Phys. B159 (1979) 99.
\item\label{kod80}
J. Kodaira, Nucl. Phys. B165 (1980) 129.
\item\label{la86}
S. G. Gorishny and S. A. Larin, Phys. Lett. B172 (1986) 109; \\
S. A. Larin and J. A. M. Vermaseren, Phys. Lett. B259 (1991) 345.
\item\label{ej74}
J. Ellis and R. Jaffe, Phys. Rev. D9 (1974) 1444.
\item\label{hei73}
R. L. Heimann, Nucl. Phys. B64 (1973) 429.
\item\label{fec88}
F. E. Close and R. G. Roberts, Phys. Rev. Letts. 60 (1988) 1471.
\item\label{herm}
HERMES Proposal, K. Coulter et al., DESY/PRC 90-1 (1990); \\
M. Veltri et al., Proc. Physics at HERA, Vol. 1, 447 (1991).
\item\label{ek93}
J. Ellis and M. Karliner, Phys. Lett. B313 (1993) 131.
\item\label{fec93}
F. E. Close and R. G. Roberts, Phys. Lett. B316 (1993) 165.
\item\label{ciof}
C. Cioffi degli Atti, talk at PANIC XIII, Perugia (1993).
\item\label{pvl93}
A. Donnachie and P. V. Landshoff, DAMTP preprint 93-23,
to appear in Zeit. Phys. C (1993)
\item\label{bt92}
S. D. Bass and A. W. Thomas, Phys. Lett. B293 (1992) 457.
\item\label{stan}
S. J. Brodsky, P. Hoyer, C. Peterson and N. Sakai,
Phys. Lett. B93 (1980) 451; \\
S. J. Brodsky, private communication (1993).
\item\label{nmc}
The NMC Collaboration, Phys. Lett. B258 (1991) 493.
\item\label{anom1}
J. S. Bell and R. Jackiw, Nuovo Cimento 51A (1969) 47.
\item\label{anom2}
S. L. Adler, Phys. Rev. 177 (1969) 2426.
\item\label{ja87}
R. L. Jaffe, Phys. Lett. B193 (1987) 101.
\item
A. V. Efremov and O. V. Teryaev, Dubna Preprint E2-88-287 (1988); \\
A. V. Efremov, J. Soffer and O. V. Teryaev, Nucl. Phys. B246 (1990) 97.
\item
G. Altarelli and G. G. Ross, Phys. Lett. B212 (1988) 391; \\
G. Altarelli and W. J. Stirling, Particle World 1 (1989) 40.
\item
R. D. Carlitz, J. C. Collins and A. H. Mueller, Phys. Lett. B214 (1988) 229.
\item
L. Mankiewicz and A. Schafer, Phys. Lett. B242 (1989) 455; \\
L. Mankiewicz, Phys. Rev. D43 (1991) 64.
\item
G. Veneziano, Mod. Phys. Lett. A4 (1989) 1605; \\
G. M. Shore and G. Veneziano, Phys. Lett. B224 (1990) 75,
	Nucl. Phys. B381 (1992) 23.
\item
R. L. Jaffe and A. Manohar, Nucl. Phys. B337 (1990) 509.
\item
M. Gluck, E. Reya and W. Vogelsang, Nucl. Phys. B329 (1990) 347.
\item
G. T. Bodwin and J. Qiu, Phys. Rev. D41 (1990) 2755.
\item
J. Ellis, M. Karliner and C. Sachrajda, Phys. Lett. B231 (1989) 497.
     (Note that the sign of the charm quark contribution to $g_1$ is incorrect
     in this reference.)
\item
S. D. Bass and A. W. Thomas, Nucl. Phys. A527 (1991) 519c; \\
S. D. Bass, N. N. Nikolaev and A. W. Thomas, Adelaide preprint
ADP-90-133/T80 (1990) \\
S. D. Bass, B. L. Ioffe, N. N. Nikolaev and A. W. Thomas,
J. Moscow Phys. Soc. 1 (1991) 317.
\item
S. Forte, Nucl. Phys. B331 (1990) 1.
\item
G. G. Ross and R. G. Roberts, Rutherford preprint RAL-90-062 (1990).
\item
A. V. Manohar, Phys. Rev. Letts. 65 (1990) 2511 and 66 (1991) 289.
\item
V. N. Gribov, Proc. SLAC Lepton Photon Symposium, World Scientific (1990) 59.
\item
S. D. Bass, Zeit. Phys. C55 (1992) 653.
\item
S. D. Bass, Cavendish preprint 93/2 (1993), to appear in Zeit. Phys. C.
\item
V. N. Gribov, Budapest preprint KFKI-1981-66 (1981).
\item
M. Shifman, Phys. Rep. 209 (1991) 341.
\item
M. A. Shifman, A. I. Vainshtein and V. I. Zakharov,
Nucl. Phys. B147 (1979) 385, 448;
\item
J. D. Bjorken and S. D. Drell, {\it Relativistic quantum fields}, McGraw Hill
      (1965).
\item
D. Amati, R. Petronzio and G. Veneziano, Nucl. Phys. B140 (1978) 54;
B146 (1978) 29; \\
S. B. Libby and G. Sterman, Phys. Rev. D18 (1978) 3252, 4737; \\
A. H. Mueller, Phys. Rev. D18 (1978) 3705; \\
R. K. Ellis, H. Georgi, M. Machacek, H. D. Politzer and G. G. Ross,
Nucl. Phys. B152 (1979) 285.
\item
V. N. Gribov and L. N. Lipatov, Sov. J. Nucl. Phys. 15 (1972) 439; \\
L. N. Lipatov, Sov. J. Nucl. Phys. 20 (1974) 181; \\
Yu. L. Dokshitzer, Sov. Phys. JETP 46 (1977) 641;
\item
S. D. Ellis, Z. Kunszt and D. E. Soper, Phys. Rev. Lett. 69 (1992) 3615.
\item
S. Catani, Yu. L. Dokshitzer and B. R. Webber, Phys. Lett. B285 (1992) 291; \\
B. R. Webber, CERN preprint TH-6871/93 (1993).
\item
R. G. Roberts, {\it The structure of the proton}, Cambridge UP (1990).
\item
J. Zinn-Justin, {\it Quantum field theory and critical phenonema},
Oxford UP (1989).
\item
P. A. M. Guichon, l'Aquila lectures, Saclay preprint DPhN-Saclay-9136 (1990).
\item
W. Bardeen, Nucl. Phys. B75 (1974) 246;
R. J. Crewther, Acta Physica Austriaca Suppl. XIX 47 (1978) 47.
\item
For a review, see
R. J. Crewther, ``Chiral Properties of QCD", in Field Theoretical Methods
in Particle Physics, Kaiserslautern 1979, ed. W. Ruhl, Vol. 55B NATO
Study Institute Series, Plenum (1980) p.529.
\item
G. 't Hooft and M. Veltman, Nucl. Phys. B44 (1972) 189.
\item
A. Brandenburg, O. Nachtmann and E. Mirkes,
Heidelberg preprint HD-THEP-93-13 (1993); \\
O. Nachtmann and A. Reiter, Zeit. Phys. C24 (1984) 283.
\item
The NA-10 Collaboration: \\
S. Falciano et al., Zeit. Phys. C31 (1986) 513; \\
M. Guanziroli et al., Zeit. Phys. C37 (188) 545.
\item
S. D. Bass and A. W. Thomas, Phys. Letts. B312 (1993) 345.
\item
M. Ahmed and G. G. Ross, Nucl. Phys. B111 (1976) 441; \\
G. Altarelli and G. Parisi, Nucl. Phys. B126 (1977) 298.
\item
C. H. Llewellyn Smith, Oxford preprint OX-89/88 (1988).
\item
J. Schwinger, Phys. Rev. 82 (1951) 664.
\item
L. L. Frankfurt et al., Phys. Lett. B320 (1989) 141.
\item
F. E. Close and R. G. Milner, Phys. Rev. D44 (1991) 3691.
\item
R. Carlitz and J. Kaur, Phys. Rev. Lett. 38 (1977) 673;
A. Schafer, Phys. Lett. B208 (1988) 175.
\item
F. E. Close and A. W. Thomas, Phys. Lett. B212 (1988) 227.
\item
A. W. Schreiber, A. W. Thomas and J. T. Londergan, Phys. Rev. D42 (1990) 2226.
\item
A. W. Schreiber, A. I. Signal and A. W. Thomas, Phys. Rev. D44 (1991) 2653.
\item
R. L. Jaffe and G. G. Ross, Phys. Lett. B93 (1980) 313; \\
R. L. Jaffe, Nucl. Phys. B229 (1983) 205.
\item
The phenomenological term is taken to be
$ -0.08 x^{0.5} (1-x)^2$ for $x\leq 0.45$, while beyond this
 it is cut-off by multiplying by $\exp (-35(x-0.45)^2)$.
\item
P. V. Landshoff and J. C. Polkinghorne, Phys. Rev. D18 (1978) 153; \\
R. M. Woloshyn, Nucl. Phys. A496 (1989) 749; \\
B. Badelek and J. Kwiecinski, Nucl. Phys. B370 (1992) 278; \\
W. Melnitchouk and A. W. Thomas, Phys. Rev. D47 (1993) 3783.
\item
For $R=0.6fm$ the phenomenological term is taken to be
$ -0.07 x^{0.4} (1-x)^3$ for $x \leq 0.45$, which
is multiplied by the large $x$ cut-off
$\exp(-5 (x-0.45)^2)$ for $x\geq 0.45$.
\item
D. H. Saxon, Rutherford preprint RAL-89-078 (1989).
\item
A. Schafer, J. Phys. G16 (1991) L121.
\item
S. Klimt, M. Lutz and W. Weise, Nucl. Phys. A516 (1990) 429; \\
U. Vogl, M. Lutz, S. Klimt and W. Weise, Nucl. Phys. A516 (1990) 469; \\
U. Vogl and W. Weise, Prog. Part. Nucl. Phys. 26 (1991) 195; \\
K. Steininger and W. Weise, Phys. Rev. D48 (1993) 1433.
\item
A. E. Dorokhov, N. I. Kochelev and Yu. A. Zubov,
Int. J. Mod. Phys. A8 (1993) 603.
\item
F. Myhrer and A. W. Thomas, Phys. Rev. D38 (1988) 1633; \\
H. Hogaasen and F. Myhrer, Z. Phys. C48 (1990) 295.
\item
H. Hogaasen and F. Myhrer, Phys. Lett. B214 (1988) 123.
\item
A. Schreiber and A. W. Thomas, Phys. Lett. B215 (1988) 141.
\item
A. Abbas, J. Phys. G15 (1989) L195.
\item
F. E. Close, Phys. Rev. Letts. 64 (1990) 361.
\item
E. M. Henley et al., Few Body Systems, Suppl. 6, (1992) 66.
\item
D. B. Kaplan and A. V. Manohar, Nucl. Phys. B310 (1988) 527.
\item
R. D. Carlitz and A. V. Manohar,
Proc. Penn State Polarized Collider Workshop, World Scientific (1990).
\item
L. A. Ahrens et al., Phys. Rev. D35 (1987) 785.
\item
E. Witten, Nucl. Phys. B104 (1976) 445.
\item
W. Melnitchouk, A. W. Schreiber and A. W. Thomas, ``Deep Inelastic
Scattering from Off-Shell Nucleons'',PSI preprint PSI-PR-93-13 (1993).
\end{enumerate}
\pagebreak

\begin{center} {\bf Figures} \end{center}
\begin{enumerate}
\item
{\bf Fig. 1: Deep Inelastic Scattering. \\}
Here, we work in the LAB frame where the proton target
(mass $M$) has momentum $p^{\mu}=(M;0_{T},0)$ and polarisation
$S^{\mu}$.
The incident muon (mass $m$) carries momentum
$ k^{\mu} = (E; \vec{k}) $ and polarisation $s^{\mu}$.
It is scattered through an angle $\theta$ so that the scattered
muon emerges with momentum $k^{' \mu} = (E^{'}; \vec{k}^{'})$.
The exchanged photon carries momentum $q^{\mu} = (k-k^{'})^{\mu}$.
We measure the total inclusive cross section so that the
final state hadrons $X$ are not separated.

\item
{\bf Fig. 2:}
The world data on $g_1^p(x)$ [7,19].

\item
{\bf Fig. 3:}
The convergence of the first moment
$\int_{x_{m}}^{1} dx \ g_{1}(x)$ as a function of $x_{m}$
in the EMC data [7].

\item
{\bf Figs. 4a-4c:}
Fig.4a shows the world data for $x g_1^p$ [7, 19] together
with the bag model expectation
[81] (solid curve), for a bag radius $R=0.8$fm.
The dashed curve is obtained by adding a phenomenological
term to fit the data.
Figs.4b and 4c show
the naive bag (solid curve) and OZI corrected (dashed curve) prediction
for $xg_1^n$ and $xg_1^d$ respectively, together with the
SLAC neutron data [22] and the SMC data for $xg_1^d$ [21].

\item
{\bf Figs. 5a-c:}
Fig.5a shows the world data for $x g_1^p$ [7,19]
together with the bag model expectation
[82] (solid line), for a bag radius $R=0.6$fm.
We fit the data by adding a phenomenological term (dashed curve).
Figs.5b and 5c show
the naive bag (solid curve) and OZI corrected (dashed curve) prediction
for $xg_1^n$ and $xg_1^d$ respectively, together with
the SLAC neutron data [22] and the SMC data for $xg_1^d$ [21].

\item
{\bf Fig. 6:}
The contribution to $g_1^p$ from $\Delta g(x)$.
We show the combination
$(g_{1}^{EMC data} + g_1^G)(x)$ (dashed curve) where
$C^g(x, \alpha_s)$ is calculated in the $\overline {MS}$ scheme.
We use the parametrisation equ.(44) with $\alpha = 0.2$
for $\Delta g(x)$.
The bold curve is the EMC fit and small x extrapolation.
Clearly, this gluon contribution is concentrated at small $x$ -
essentially outside the range of the present data.

\item
{\bf Fig. 7:}
Contribution to the neutrino proton cross section where
the $Z^{0}$ couples to the proton via two gluons.
\end{enumerate}
\end{document}